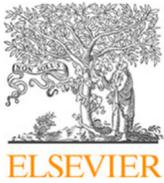
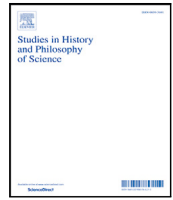

# Blindspots of empiricism in the discovery of chaos theory

Brett Park

*Department of History and Philosophy of Science, University of Pittsburgh, United States of America*

A B S T R A C T

Chaos theory is a branch of classical physics, founded in the 1960s-70s, that studies systems whose solutions are sensitively dependent on their initial conditions. For many, it is surprising that chaos theory arrived so late. However, through the work of Henri Poincaré, we know that much of the math of chaos was understood by some 70 years prior. Furthermore, through the writings of Poincaré's colleagues — Jacques Hadamard and Pierre Duhem — we also see a detailed understanding of the chaos found in his work. They also have explicit reasons of why the math of chaos was to be ignored. It was a strict form of empiricism — positivism — causing them to label chaos as "useless" and "meaningless" mathematics because it was thought to be ungrounded in experience. In this paper, I describe how the empiricist tenets of positivism exiled chaos from physics following Poincaré.

> The day will perhaps come when physicists will no longer concern themselves with questions which are inaccessible to positive methods, and will leave them to the metaphysicians. That day has not yet come; man does not so easily resign himself to remaining for ever ignorant of the causes of things.
>
> Poincaré (1905, 223).

## 1. Introduction

Chaos theory is the study of classical systems that are sensitively dependent on their initial conditions. When the field took off in the 1960s–70s, it was hailed as the third scientific revolution of physics of the 20th century, following relativity and quantum mechanics (Gleick, 1987, 33). However, unlike quantum mechanics and relativity, chaos theory does not contradict any of the fundamental posits of classical physics, a 300 year old theory. Rather, it overturned physicists' assumptions about what types of behavior is generic in classical systems — e.g. replacing stability/periodicity with instability/aperiodicity. This has led many researches to wonder over why these ideas appear so late. This is particularly puzzling when combined with another well-known fact; in Henri Poincaré's 1890 work on the three-body problem, many of the central ideas of chaos were recognized and developed by the influential physicist 70 years before they gained wider recognition. Poincaré's work would be largely forgotten in the intervening years, until it was picked up by topologist Stephen Smale in the 1960s.

In this paper, I investigate why Poincaré's early insights into chaos seemingly vanished from physics shortly after they appeared, presenting a novel history of this period. The influence of a strict empiricist philosophy — positivism — led researchers to reject chaos as "meaningless" and "useless" mathematics. To show this, I examine the writings of people near Poincaré during this time, focusing particularly on the geometer Jacques Hadamard and physicist/philosopher Pierre Duhem. In them, we find clear recognition of some of the essential mathematical aspects of chaos theory that Poincaré had uncovered. They also give lucid positivist critiques of why these results lacked any physical significance. I trace the conflict back to two central ideas of positivism. First, positivists rejected the notion of unobservable causes. Second, they understood the laws of nature as mere summaries of observed regularities. In chaotic systems, small, unobservable differences in initial conditions can create large, observable ones in the future. Furthermore, chaotic systems alerts us to the fact that law-governed, deterministic dynamics can give rise to a lack of observable correlations. Both of these ideas proved unintelligible from a positivist philosophy. Throughout this period, we repeatedly see these key figures reject chaos' physical significance on these grounds.

The body of the paper is divided into two sections, Sections 2 and 3. In Section 2, I describe how Poincaré's developed his insights into chaos, and how these insights were neglected shortly thereafter. I give a brief description of chaos in Section 2.1. In Section 2.2, we see how Henri Poincaré uncovered many of the mathematical ingredients of chaos over half a century before the chaos "revolution". In Section 2.3, I describe how Poincaré's results would be reintroduced into physics via Stephen Smale. In Section 2.4, I detail how other authors have tried to account for the anomalous gap between Poincaré and Smale.

In Section 3, I offer a novel account for why chaos was neglected following Poincaré. The influence of positivism led to chaos being banished from what was considered physics' proper domain. In Section 3.1, I detail the roots and growth of positivism up until the time of Poincaré. I describe the work of Jacques Hadamard in Section 3.2, which exhibits both a solid understanding of chaos and a positivist grudge against its meaningfulness. Hadamard's rationale is greatly extended in Section 3.3 by his friend Pierre Duhem. In Section 3.4, we explore the broader acceptance of the positivist critique within the physics community. I offer some remarks on the role of philosophy of science in shaping scientific research in Section 3.5. I conclude in Section 4 by integrating these ideas into the existing history.

## 2. The mathematical lineage of Chaos

### 2.1. What is Chaos?

To frame the discussion, we should give a brief description of chaos. A classical system with $N$ variables can be modeled in an $N$-dimensional space called a *phase space*. A point in phase space represents the state of the system at a given time. The state of the system evolves according to a set of differential equations, where the solution to an initial value problem will trace out a curve in phase space called a *trajectory*. Call a trajectory *periodic* if it returns to the same point in phase space, *aperiodic* otherwise. In chaotic systems, almost all trajectories confined in the chaotic region will be aperiodic.

The most widely agreed on modern definition of chaos includes the following three properties:

(1) Sensitive dependence on initial conditions,
(2) The existence of at least one dense trajectory,
(3) Density of periodic trajectories.

Sensitive dependence, often considered the hallmark of chaos, means that two trajectories which start out close together end up far apart. "Dense" in the sense of (2) and (3) mean that in a given phase space, a certain set either contains every point in that phase space or contains a point that is arbitrarily close to that point. In this context, (2) means that there is one trajectory that gets arbitrarily close to every other point in phase space. (3) means that there is a periodic orbit that passes arbitrarily close to any point in phase space.

In the rest of this section, I provide a brief description of the mathematical lineage of chaos theory, focusing on Henri Poincareé's pioneering 1890 results on the three body problem and their reintroduction by Stephen Smale in the 1960s.

### 2.2. Henri Poincaré on the three-body problem

One of the oldest problems in classical mechanics is the stability of the solar system; will the stable, quasi-periodic behavior we observe in the solar system continue indefinitely or will the mutual interactions of the planets eventually lead to instability? Newton believed that his dynamics would entail instability, causing him to hypothesize that God had to periodically intervene to stabilize the orbits (1704, 378). However, it was not until after Henri Poincaré's famous 1890 essay on the three-body problem, as well as his subsequent three-volume treatise on celestial mechanics, that the prospect of instability seemed assured.[1] It is also in these works that we find the math of chaos the first time. We will focus on those results.[2]

Throughout his work, Poincaré would employ a qualitative approach to studying differential equations. Instead of looking for a general solution to the equations, Poincaré would find specific, periodic solutions and analyze the situation around them. This would allow Poincaré to deduce much about the behavior of the system without having the explicit solution in hand.

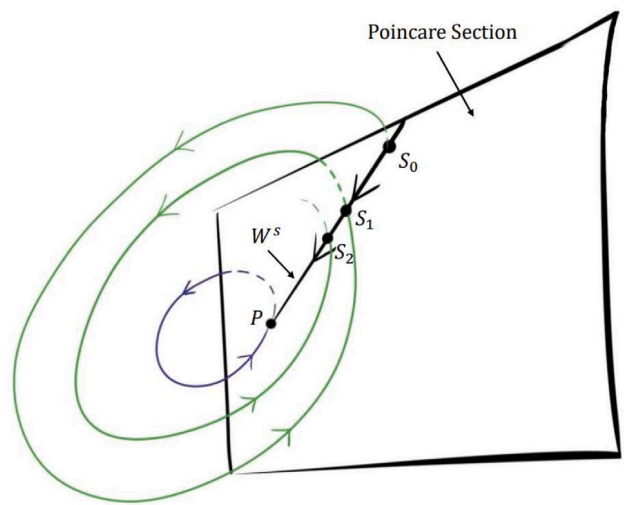

**Fig. 1.** Poincaré section of a limit cycle trajectory (blue), creating the fixed point $P$. The green line is a trajectory caught on the stable manifold $W^S$ that will asymptotically approximate the limit cycle as $t \to \infty$. Accordingly, intersection points $S_n$ are mapped asymptotically closer to $P$ along $W^S$ within the Poincaré section. (For interpretation of the references to color in this figure legend, the reader is referred to the web version of this article.)

Poincaré justified the examination of periodic solutions in the following manner (1967a, 63). He recognized that the initial conditions leading to periodic trajectories would be probability zero over phase space. However, he conjectured that periodic orbits would be densely packed over phase space. This means that any aperiodic initial condition will be approached arbitrarily closely by a periodic initial condition. Thus, there will always be a periodic trajectory that can serve as a good first approximation for any aperiodic trajectory.

Poincaré's clearest depiction of chaos comes in his examination of a special type of periodic trajectory: saddle limit cycles. A limit cycle is a periodic trajectory that is asymptotically approached by other trajectories. In the simple case, if we cut out a section of phase space that is transverse to the limit cycle (Poincaré section) and map where the limit cycle intersects with the section, the cycle is mapped to the same point under the dynamics (Fig. 1). Within the phase space, there will be one or more manifolds where any trajectory on that manifold will asymptotically approach the limit cycle. Manifolds where trajectories asymptotically approach the limit cycle as $t \to \infty$ are called stable manifolds $W^S$, whereas unstable manifolds $W^u$ have trajectories that approach the limit cycle as $t \to -\infty$. The Poincaré section will likewise intersect with these manifolds. The result of such an intersection is shown as $W^S$ in Fig. 1. A saddle limit cycle has both a stable and an unstable manifold. In its Poincaré section, it will appear as a saddle point $P$ at the center of the two manifolds. It is possible for these two manifolds to intersect within the Poincaré section, and the point at this intersection will be mapped asymptotically closer to $P$ as $t \to \pm\infty$. Poincaré called trajectories that pass through such a point "homoclinic" (1967c, 377).[3] Poincaré proved that so long as these manifolds intersect once within this Poincaré section, they will intersect an infinite number of times in that section. Thus, one homoclinic point implies an infinite number of them. This results in an extremely convoluted structure that Poincaré describes as a

> grid with infinitely serrated mesh. Neither of the two curves must ever cut across itself again, but it must bend back upon itself in a very complex manner in order to cut across all of the meshes in the grid an infinite number of times. The complexity of this figure will be striking, and I shall not even try to draw it. Nothing is more suitable for providing us with an idea of the complex nature of the three-body problem, and of all the problems of dynamics in general, where there is no uniform integral (1957, 381–382).

---

[1] English translations of his 1890 essay and three-volume treatise can be found in Poincaré (1967a, 1967b, 1967c, 2017).

[2] For a detailed account of Poincaré's work on the three-body problem, see Barrow-Green (1997).

[3] In his 1890 essay, he calls these trajectories "doubly asymptotic" (Poincaré, 2017, 200).

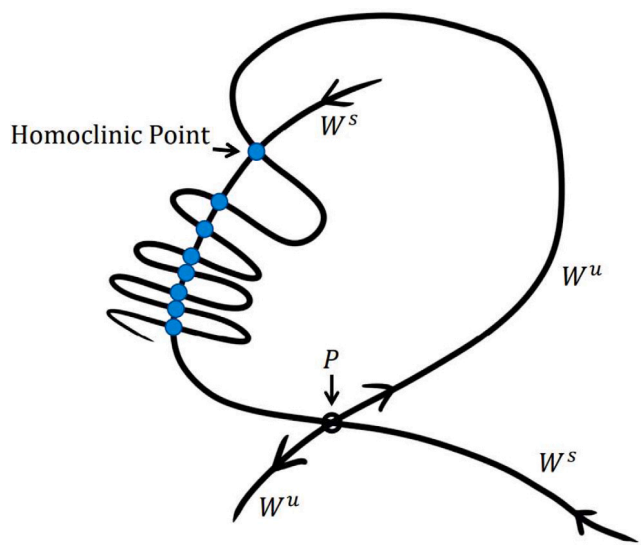

**Fig. 2.** The beginning of a homoclinic tangle within a Poincaré section.
*Source:* Adapted from Hilborn (2000, 141).

The diagram described is shown in Fig. 2, and has been dubbed the "homoclinic tangle". The lines represent a Poincaré section of the stable $W^s$ and unstable $W^u$ manifolds, not individual trajectories. A point on a manifold will be mapped to another point on that manifold. A homoclinic trajectory (shown in blue) passes through a homoclinic point where the stable and unstable manifolds intersect. Since the trajectory must remain on both manifolds, a single homoclinic point entails an infinite number of intersections.

From this situation, we can tell a lot about the behavior of the system. Points near the stable manifold will be pulled into the saddle point $P$, and points near the unstable manifold will be repelled. Regions of initial conditions near $P$ are stretched and folded back onto themselves under time-evolution, with trajectories that start nearby ending up far apart. In a letter to Heinrich Hertz, Poincaré says that the complexity of the solution around the homoclinic points can be understood by "seeing the orbits we might trace become tangled with each other in a more and more complicated way" (Poincaré, 1890). This complex tangle of trajectories is how chaotic systems exhibit sensitive dependence while keeping trajectories confined within a bounded region.

It is disputed how squarely Poincaré's work on homoclinic orbits fits within the modern definition of "chaos" described earlier.[4] However, it is not disputed that Poincaré's work would provide the launching off point for many of the important advances in the field to come. Anyone who has read a textbook or taken a course on chaos can attest to the volume of techniques and results that bear his name. Through his writings, we can see that Poincaré clearly recognized the importance of homoclinic points in understanding non-integrable classical systems, which he saw as comprising most classical systems.[5] This is why Robert Hilborn writes that "Even a cursory reading of the history of chaos… shows that Poincaré knew about, at least in a rough way, most of the crucial ideas of nonlinear dynamics and chaos" (2000, 62). We are now in a position to understand what this means. In Poincaré's work, we find descriptions or conjectures of the following:

- Sensitive dependence on initial conditions,
- Periodic trajectories that are measure zero but densely packed over phase space,
- That the above behavior is widespread in classical systems.

---

[4] To see why Poincaré's work on the three body problem does not fit modern definitions of chaos theory, see Zuchowski (2014).
[5] For an example of Poincaré making the latter claim, see Poincaré (1914, 36).

It would be another 70 years before these ideas began to be systematized by the physics community under the field we now call "chaos theory".

### 2.3. Stephen Smale

Somehow, following Poincaré the trail goes cold. These results would be largely neglected, with a few scattered exceptions such as the mathematician Birkhoff (1913, 1931).[6] When the long overdue reintroduction of Poincaré's work to physics finally came, it came by way of pure mathematics. It would ultimately be topologist Stephen Smale who would uncover the significance of homoclinic points.

Smale reflects on how he found out about these results:

Unfortunately, the scientific community soon lost track of the important ideas surrounding the homoclinic points of Poincaré. In the conferences in differential equations and dynamics that I attended in the late 50 s, there was no awareness of this work. Even [Norman] Levinson never showed in his book, papers, or correspondence with me that he was aware of homoclinic points. It is astounding how important scientific ideas can get lost, even when they are aired by leading mathematicians of the preceding decades. I learned about homoclinic points and Poincare's work from browsing in Birkhoff's collected works (1998, 44)

Smale's chance encounter with Poincaré's homoclinic points marked an inflection point in the mathematical study of chaos.

Throughout the 1960s, Smale would greatly expand our understanding of the homoclinic tangle. Smale realized that we could abstract away from specific dynamical systems by focusing purely on the topological features of homoclinic tangles (Smale, 1967). He distilled these features down to a simple map from the unit square onto itself called the horseshoe map, which stretches the unit square and then folds it back onto itself. He then proved that the horseshoe map acting on its invariant (Cantor) set is homeomorphic to the shift map acting on the set of infinite binary sequences (The Birkhoff–Smale Theorem). This latter map is provably chaotic, in the sense of (1)–(3) above.

Smale's work marked a boon for the mathematical analysis of chaos, as it was quickly found that most interesting chaotic behaviors could likewise be rigorously studied using simple maps. In 1961, Smale would give a lecture on his horseshoe map in Kiev, attended by future luminaries in the field such as Yakov Sinai and Dmitri Anosov. After learning of Smale's horseshoe, Anosov recalls "At that moment the world turned upside down for me, and a new life began" (2006, 10). A similar sentiment soon swept across the field.

### 2.4. A puzzling gap

Many authors have noticed this strange state of affairs. Historian of mathematics June Barrow-Green writes that "It is conspicuous that during the early years of the 20th century no serious attempt was made to investigate further the behavior of Poincaré's doubly asymptotic solutions" (1997, 219). Likewise, David Ruelle calls this a "puzzling historical gap" (1993, 49) and Stephen Kellert has dubbed this problem "the nontreatment of chaos" (1993, 120). Many explanations have been offered.

---

[6] Other notable examples include the work of Arnold (1963), Kolmogorov (1954), and Moser (1962) in what is now known as the KAM theorem.

One of the most prominent explanations offered is that investigators lacked the computing resources which have proven to be instrumental in exploring chaos. Because chaotic systems are nonlinear, they typically do not have closed-formed solutions. Digital computers can be used to numerically approximate solutions to nonlinear systems. Barrow-Green writes that the puzzle can be "To a large extent… explained" by the lack of adequate computing technology (1997, 219). Similar statements have been frequently made in the literature.[7] According to Douglas Hofstadter, the computer played the same role in the study of chaos theory that the telescope and particle accelerator would play in other areas of physics (1985, 365). Robert Hilborn makes the point that "if Poincare had had a Macintosh or IBM personal computer, then the field of nonlinear dynamics would be much further along in its development than it is today" (2000, 62). These authors are no doubt correct that digital computers played a large role in the blossoming of chaos theory as a field of inquiry during the 1960s–70s.

However, this explanation is by itself insufficient. We have already seen that, as a matter of history, the reintroduction of Poincaré's work came not by way of numerics but pure mathematics via Stephen Smale. Smale's student, Morris Hirsch, would recount that Smale's results were almost entirely obtained by way of pure math, with computers playing little role in this period of discovery (Hirsch, 1989, 7). Why did Poincaré's insights die out? To use the analogy of Hofstadter, in particle physics the Higgs boson was expected for a long time on theoretical grounds before we had the technology to empirically confirm it. Likewise, it is plausible that many of the behaviors found in chaotic systems should have been expected on theoretical grounds. But when they finally did have digital computers, physicists bumped into chaos as if by accident, as if they were not expecting it to be there (Gleick, 1987, Ch. 1).

Other auxiliary explanations have been proposed. As Kellert (1993, 158) and Ruelle (1993, 49–50) point out, the advent of quantum mechanics (and relativity) meant that the energy of physicists had largely been diverted away from classical physics at this time. There were also political forces at play. The two world wars would occupy and reshape the physics community in various ways. After the second world war, some of the most important work in nonlinear dynamics happened behind the Iron Curtain in the Soviet Union (Aubin & Dalmedico, 2002).[8] A full accounting of why chaos theory was not further developed during this period is complicated, and each of the explanations bears merit.

However, there is an ahistorical reading of this period that could be drawn from these explanations alone. It is not the case that after Poincaré, the nature of chaos was initially recognized widely across physics, only for further research to be put on hold while attention was diverted elsewhere. What we do find is the mathematical aspects of Poincaré's results recognized primarily by those working near to him. However, these figures would mislabel chaos as physically uninteresting. After this, even the mathematical ideas are all but forgotten. In the rest of the paper, I will examine why the early ideas were initially exiled from physics and subsequently forgotten.

## 3. Positivism and Chaos theory

In the last section, I have described how Poincaré's work on the three body problem planted the early seeds for chaos research. Many cogent explanations have been offered for why these seeds took so long to sprout. My aim with the rest of this paper is to fill in a detail in this history that has been overlooked. By zooming in on figures surrounding Poincaré, we find a surprisingly good understanding of the mathematical nature of chaos introduced by his work. However, in their reactions, we find a certain set of empiricist commitments explicitly defining the mathematical results as "meaningless" and "useless" for physics. These commitments can be traced back to the positivist philosophy of science that had become endemic at the time. Thus, we can see an over-zealous strand of empiricism blunting the impact of Poincaré's work with the very people who should have been its champions.

### 3.1. An era of positivism

The goal of positivist philosophy was to rid the sciences of idle metaphysical speculation. Here, I will detail how positivist attitudes came to pervade European science by the time of Poincaré, spelling out which ideas of positivism would come into conflict with chaos research. In following sections, I will give detailed cases where this conflict plays out.

The earliest roots of positivism go back to David Hume. Hume was rabidly anti-metaphysics, declaring that any philosophical term that cannot be traced back to sense experience is "employed without any meaning or idea" (1748, 15). Hume was an early inspiration in positivists' attempt to root out metaphysics from science. When 'Positivism' was officially christened by Auguste Comte in the mid 1800s, Comte would write that "Hume constitutes my principal philosophical precursor" (1852/1909, 5).

Comte historicized Hume's critique of metaphysics. Compte believed that our understanding of nature progresses through three stages: the theological stage, the metaphysical stage, and finally the positive stage (Comte, 1853). During the theological stage, humans would give and accept explanations of phenomena in terms of divine or supernatural entities. This would give way to the metaphysical stage, where people would no longer appeal to supernatural entities but to unobserved abstract powers or forces. Lastly, in the positive stage, unobserved causes would be abandoned, leaving only the scientific laws which serve to organize and classify experience. This rejection of unobservable causes and understanding of laws is central to positivist philosophy and will play a large part in the initial rejection of chaos.

Positivism became hegemonic throughout European science after Comte. John Stuart Mill, a lifelong admirer and frequent correspondent of Comte, writes that

> For some time much has been said, in England and on the Continent, concerning "Positivism" and "the Positive Philosophy". Those phrases… have emerged from the depths and manifested themselves on the surface of the philosophy of our age… They are symbols of a recognized mode of thought, and one of sufficient importance to induce almost all who now discuss the great problems of philosophy … to take what is termed the Positivist view of things into serious consideration, and define their own position, more or less friendly or hostile, in regard to it (1866, 1–2).

To understand the intellectual milieu at the time of Poincaré, Mill's last point is essential. Even the "anti-positivist" schools of philosophy such as the French spiritualists — including Émile Boutroux (Poincaré's brother-in-law) and Félix Ravaisson — defined their philosophies from within a positivist frame. That is, they tried to make room for other types of truth, such as spiritual truth, amid the backdrop of positivist thought. Thus, the influence of positivism had spread far beyond those who explicitly accepted the title and its association with Comte and secularism.

Following Hume and Comte, positivist philosophy would routinely see the unobservable labeled as "meaningless". A well known example is the rejection of the notion of absolute space. Ernst Mach, an influential late 19th century physicist and philosopher, writes of a hypothetical test of Newton's bucket experiment: "the experiment is impossible, the idea is meaningless, for the two cases are not, in sense-perception, distinguishable from each other. I accordingly regard these two cases as the same case" (1893, 512). On this point, Poincaré concurred with Mach when he writes that "Whoever speaks of absolute space uses a word devoid of meaning" (Poincaré, 1914, 93). Mach also famously rejected the existence of atoms and molecules due to their unobservable nature (see Blackmore, 1972, 204–231). These terms were unobserved, metaphysical causes to be abandoned in the progression of science.

---

[7] Mathematician John Franks writes that the ubiquity of chaos "is a surprising fact… because it was invisible before the computer, but with computers it is easy to see, even hard to avoid" (1989, 66).

[8] Philosophical explanations of this puzzle tend to focus on how chaos subverted the traditional "clockwork" picture of Newtonian mechanics (Peterson, 1993; Kellert, 1993, 119; Barrow-Green, 1997, 219–220; Gleick, 1987, 12).

Positivists believed science's objective was to ascertain the laws connecting empirical phenomena. Comte viewed laws of nature as mere descriptions of the most regular associations of experience, not the causes underlying experience. Mach forcefully advanced and popularized this idea. In *The Science of Mechanics*, he writes "the principles of science are all abstractions that presuppose repetitions of similar cases" (1893, 504). In a later work, he writes that

> Everything that we can want to know is given by the solution of a problem in mathematical form, by the ascertainment of the functional dependence of the sensational elements on one another. This knowledge exhausts the knowledge of "reality" (1914, 369).

For Mach, the goal of science is to provide economy of thought, a way of storing the regularities of experience. The only reason we can bring order to our flux of sensations is because memory allows us to relate what we are observing now to things we have observed in the past. Accordingly, we build up an association of which similar events are followed by which others. In this way, laws of nature are the ultimate memory saving device; they encode phenomenological associations in compact mathematical form.

This rough conception of laws, and their role in scientific investigation, was widely circulated by some of the most influential scientists of this period. James Clerk Maxwell writes that "Physical science is that department of knowledge which relates to the order of nature, or, in other words, to the regular succession of events" (1877/1920, 1). In *Science and Hypothesis*, Poincaré asserts

> The object of mathematical theories is not to reveal to us the real nature of things; that would be an unreasonable claim. Their only object is to co-ordinate the physical laws with which physical experiment makes us acquainted, the enunciation of which, without the aid of mathematics, we should be unable to effect (Poincaré, 1905).

Elsewhere, we see Poincaré rehearsing Mach's ideas of science as economy of thought.[9] In the social sciences, Émile Durkheim would found the discipline of sociology on the premise that social science should seek out general laws that systematize observations (1895).

Why should this empiricist philosophy of science have any bearing on research into chaos? The first problem comes from labeling unobservable causes as "meaningless" metaphysics. For positivists, the following sort of claim is to be rejected:

> Something we can observe (meaningful) is caused by something we cannot observe (meaningless).

The mathematics of chaos theory suggests that even imperceptible differences can rapidly become large differences. But this leads to the claim that observable future facts can be "caused" by unobservable facts about the present. This proved to be unintelligible from a positivist philosophy of science. As we will see, questions whose answers depend sensitively on initial conditions are repeatedly labeled "meaningless".

Secondly, chaos did not conform with the positivist conception of laws. The long-term dynamics of chaotic systems cannot be understood in terms of correlations of similar sense data. In fact, one of the interesting properties of chaotic systems is that, given enough time, there will be no detectable correlations between past and future states (see Werndl (2009)). Even before Poincaré, Maxwell struggled to fit chaos into this tidy scheme of law:

> it is only in so far as stability subsists that principles of natural law can be formulated... In so far as the weather may be due to an unlimited assemblage of local instabilities, it may not be amenable to a finite scheme of law at all (1877/1920, 13–14).

In this surprisingly modern description of how chaos manifests in the weather, Maxwell places instability (i.e. chaos) outside the boundary of "natural law".[10] Comte had also expressed doubts about the prospects of a science of meteorology, stating that the laws of the weather "are still, and perhaps, always will be, unknown" (1865, 26). Comte blames the "multiplicity" of the relevant laws needed to obtain meteorological predictions and "Our feeble faculties" of deduction, rather than the unintelligible idea that the laws themselves give rise to a lack of observable correlations (1853, 47).

### 3.2. Jacques Hadamard

We have just seen how some central ideas of positivism obscure some key insights into chaos. In two of Poincaré's French contemporaries, Jacques Hadamard and Pierre Duhem, the ideas of positivism and chaos collide head on. These cases are astounding because both Hadamard and Duhem provide nuanced descriptions of the math of chaos, before explicitly rejecting it as — variously — "meaningless", "useless", or not "physical". We will begin with the work of Hadamard.

Jacques Hadamard was a prominent French mathematician who worked alongside Poincaré. Due to his explorations of geodesic deviation in spaces with negative curvature, Hadamard is sometimes co-credited with Poincaré as being one of the first to discover chaos in dynamical systems (Steiner, 1994). However, in his writings we see ideas from positivism driving Hadamard to reject the physical significance of these results.

By the time he began work on the mathematics of chaos, Hadamard had frequent exposure to positivist philosophy. He was a lifelong friend of avowed positivist Pierre Duhem, starting from their days as students at École Normale Supérieure.[11] Hadamard would favorably reference Duhem's philosophy of science throughout his life.[12] A second major positivist influence was Émile Durkheim, whom he became colleagues with when he moved to Bordeaux in 1893 (Maz'ya & Shaposhnikova, 1999, 61). Hadamard would also regularly attend prominent philosophical meetings with the likes of Poincaré, Bertrand Russell, and Henri Bergson, where the ideas of positivism were frequently connected to contemporary science.[13]

In his 1898 paper on spaces of negative curvature, Hadamard gives a candid description of the math of chaos. Hadamard adopts Poincaré's qualitative approach to studying differential equations. The trajectory of a frictionless particle traveling in these spaces is equivalent to

---

[9] Poincaré writes "The celebrated Viennese philosopher Mach has said that the role of science is to produce economy of thought, just as machines produce economy of effort. And that is very true" (Poincaré, 1914, 28).

[10] Poincaré also connects sensitive dependence to the weather (1914, 68).

[11] Hadamard (1927, 467) wrote of their early school days:

> In these long and precious conversations during which, from the moment of my entry to the École, our friendship grew, how I felt him being thrilled by the genius of Hermite, or that of Poincaré, whose works he followed better than most of us could (I mean the most specialized in mathematics)!

[12] See his remarks in Duhem (1954, 216) and Hadamard (1927, 469).

[13] See the proceeds from First International Congress of Philosophy (1900), as well as Maz'ya and Shaposhnikova (1999, 142).

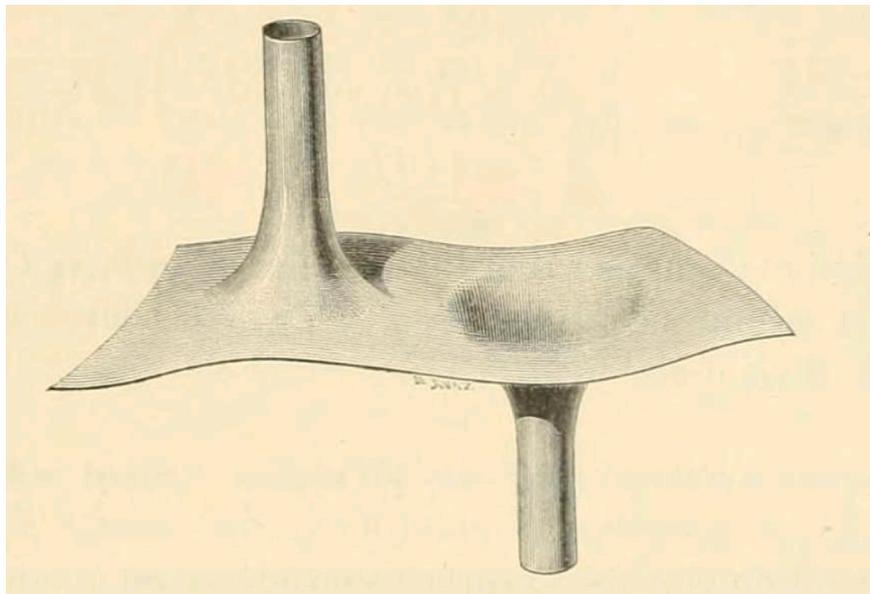

**Fig. 3.** Hadamard's system of negative curvature, found in Hadamard (1898, 39).

a geodesic in that space, the locally length-minimizing path. Thus, only the position and direction of motion is needed to determine the trajectory. Hadamard considered three dimensional embeddings of two-dimensional surfaces which would produce curvature that is everywhere negative. He gives a sample construction: take two fixed points, $P_1$ and $P_1'$ in the $(x, y)$ plane, where $\delta_1$ and $\delta_1'$ are the distances between the point $(x, y)$ and these fixed points, respectively. An embedded 2d surface of negative curvature is given by

$$z = k \log \frac{\delta_1}{\delta_1'}$$

which generates a figure with a horn extending to $z = -\infty$ around $P_1$ and a horn extending to $z = +\infty$ around $P_1'$ (Fig. 3). We can create as many horns as we like with fixed points $P_1, \ldots, P_m, P_1', \ldots, P_n'$ and an equation of the form:

$$z = k \log \frac{\delta_1, \ldots, \delta_m}{\delta_1', \ldots, \delta_n'}$$

which will yield $m + n$ horns; $m$ for $z = -\infty$ and $n$ for $z = +\infty$.

For such a system, Hadamard reasoned that if we extended each geodesic line to infinity, there would be three types of geodesics:

1. **Unbounded Trajectories**: geodesics that go off to infinity
2. **Periodic Trajectories**: closed geodesics (or geodesics which asymptotically approached a closed geodesic)
3. **Aperiodic Trajectories**: geodesics which would switch between neighborhoods of closed geodesics

Both periodic and aperiodic trajectories would loop around the horns in some bounded region. Hadamard proved that aperiodic trajectories would be approached arbitrarily closely by periodic ones, and that an infinitesimal change in either a periodic or aperiodic trajectory could lead to it switching types (1898, 66–67). He reasons that nearby trajectories will diverge exponentially. Once again, we see two of the central mathematical hallmarks of chaos — dense periodic orbits and sensitive dependence — loosely recognized well before they became widely known to the physics community.

Equally interesting as his subtle mathematical understanding of chaos is his assertion of its physical *meaninglessness*. At the end of the same paper, Hadamard relates his own work to Poincaré's work on the stability of the solar system. Hadamard points out that in astronomical problems… the constants that define the motion are given physically, that is, with errors whose amplitude reduces as the power of our means of observation increases, but which cannot be canceled out. If we only follow the trajectories for a specific time, any time, we can imagine that the errors in the initial data have been made minimal enough not to significantly alter the form of these trajectories during the aforementioned time interval. The above shows us that it is in no way legitimate to draw a conclusion from this… This may well depend… on discontinuous arithmetic properties of the integration constants (1898, 71–72).

Recall the stability of the solar system has to do with its ultimate fate; will it forever remain in the same approximate quasi-periodic configuration or will, for example, Mars be ejected towards spatial infinity similar to one of Hadamard's unbounded trajectories? What Hadamard suspects from his and Poincaré's work is that, according to the mathematical representation of the system, the ultimate answer is going to depend on the precise mathematical initial conditions, not the "physical" initial conditions which are limited by observational error. In Hadamard's system, arbitrarily small differences in initial conditions can change the trajectory to any of types 1–3. Similarly, the observable (i.e. meaningful) fact about whether Mars is ejected might depend on in-principle unobservable (i.e. meaningless) facts about the initial conditions. But we cannot appeal to unobservable causes to explain the observable! For this reason, Hadamard suspects that the question of the stability of the solar system will "cease to have any meaning" (ibid. 71).[14]

A second place we see positivism guide Hadamard's ideas on chaos is through his thoughts on "well-posed" problems in physics. Hadamard held that an initial value problem is well posed if and only if a solution (I) exists, (II) is unique, and (III) is continuously dependent on the initial conditions. A problem that fails any of these conditions is "ill-posed". Using the standard mathematical sense of continuity, (III) says that a convergence in the initial conditions will lead to a convergence of solutions. For chaotic systems of ordinary differential equations, solutions will always be continuously dependent on initial conditions. However, chaotic systems of partial differential equations can fail to satisfy (III), and thus can be "ill-posed" in the sense of Hadamard (Lorenz, 1969; Palmer et al., 2014).

---

[14] In the original French: "cesserait d'avoir un sens" (Hadamard, 1898, 71).

Hadamard's motivation for this criteria is an empiricist one. When explaining the criteria in his 1923 lectures on Cauchy problems, he explains that cases where solutions are sensitive to unobservable differences cannot "correspond to physical questions" (Hadamard, 1923, 38). Hadamard explains that any system displaying this type of sensitive dependence "would appear to us as being governed by pure chance (which, since Poincaré, has been known to consist precisely in such a discontinuity in determinism) and not obeying any law whatever" (ibid, 38). It should also be noted that the antiquated, empiricist sense of 'determinism' Hadamard uses means that an observable initial state will be associated with a unique, observable future state, rather than simply uniqueness of mathematical solutions. This definition rules chaotic systems as indeterministic precisely because they are sensitive to unobservably small variations in initial conditions. In the first meeting of the International Congress of Philosophy in 1900, attended by Poincaré, Hadamard claims that alleged instances of indeterminism are "mathematically conceivable but physically impossible" and that such cases of "indeterminacy are merely cases of unstable equilibrium, that is, cases where a small variation in the cause (of the data) produces a very large variation in the effect" (1900, 543). Here, "data" refers to the mathematical initial conditions, not observational data. This quote happens in a larger conversation about indeterminism in classical physics and celestial mechanics, in which discussants distinguish between mathematical solutions and "real, empirical" ones (1900, 544). This is a recurring theme throughout this period; sensitive dependence on unobservable differences is thought of as a mathematical pathology which can be found in certain abstract Newtonian systems but not in reality.

Hadamard's criteria for well-posed problems evolved over the years. In early works, he only included (I) and (II), while in later work he came to emphasize (III).[15] However, even in an early 1901 discussion, Hadamard appears unsure about fate of his systems of negative curvature as well as many-bodied gravitational systems, and repeats his earlier worry:

> It will above all be necessary to admit that the shape of the trajectories can depend on discontinuous, arithmetic properties of the constants of integration. Secondly, and as a consequence, important problems of Mechanics, such as that of the stability of the solar system, may fall into the category of ill-posed problems. If, in fact, we substitute the question of the stability of the solar system with the analogous question relating to the geodesics of the surfaces of which we have spoken, we find that any stable trajectory can be transformed, by an infinitesimal change in the initial data, into a completely unstable trajectory which extends to infinity, or more generally, into a trajectory exhibiting any of the forms listed in the general discussion: for example, into a trajectory asymptotic to any closed geodesic. Now, in astronomical problems, the initial data are only known physically, that is to say only within a certain degree of error. However small, this error could cause a total and absolute perturbation in the desired result (1901, 14).

Again we see Hadamard as being exquisitely attentive to the distinction between mathematics and observation. Because arbitrarily small differences in initial conditions might lead to a qualitatively different answers, Hadamard wonders whether the question about the stability of the solar system is "ill-posed".

Throughout his work, Hadamard is very concerned with the empirical grounds of physics, which he takes to be a correspondence between observable initial and final states. He is also quick to discard any mathematics that cannot be observationally grounded in this way as "meaningless" or not "physical". Thus, he repeatedly labels his own work on sensitive dependence as being, in various ways, irrelevant to our understanding of the physical world. This line of reasoning would be made even more explicit by Hadamard's friend, Pierre Duhem.

### 3.3. Pierre Duhem

Pierre Duhem was a physicist and philosopher of science. Poincaré would serve on his dissertation committee at École Normale Supérieure, (Ariew, 2022). Spurred by Hadamard and Poincaré's work, Duhem would present a detailed positivist critique of chaos in his 1906 book, *The Aim and Structure of Physical Theory*, a work he boasts as "positivist in its conclusions as well as in its origins" (Duhem, 1954, 275). The goal of Duhem's book is to describe the relation between theory and observation in physics. Today, Duhem is primarily known in philosophy of science as being one-half of the Duhem–Quine thesis i.e. that our scientific theories are underdetermined by the data. Less known is his description of "useless mathematics" and its relevance for chaos.[16]

Citing Mach as an inspiration, Duhem warns against the overreach of mathematics beyond experience. Duhem argues that physics is uniquely poised to "unite the certainties of common-sense findings with the clarity of mathematical deduction" (1954, 267). Common sense is provided by observation, and it will suggest new ways to modify the logical structure of theories. The consequences of those modifications are necessitated by mathematics. However, Duhem identifies a "mixed-zone", where these two means of certainty operate in rivalry (ibid., 267). Here, Duhem contends that mathematicians often overextend the reach of mathematics beyond its proper boundary by speculating on structure that goes too far beyond experience. He cites Mach in calling this a tendency towards "false rigor" (Duhem, 1954, 268; Mach, 1893; 82).

In an effort to police this "mixed zone", Duhem mirrors Hadamard's idea that systems with sensitive dependence lie beyond the scope of physics. First he describes a positivist conception of laws. In experiments, what we observe is a regular transition from one set of similar states to another set of similar states. Differential equations can provide a mathematical description of this transition. For Duhem, the observed states are "practical facts" and their mathematical representations are "theoretical facts" (ibid., 134). Computing a system's future state given some initial condition is a "mathematical deduction" of one theoretical fact from another (ibid., 141). On its surface, this appears to be a one-to-one map between a physical and mathematical system.

However, Duhem points out that while our mathematical representations of states take on infinite precision, such as a point in phase space, our knowledge of physical states are only ever approximate. This means that a single practical fact, obtained through observation, can always be translated into "a whole bundle of theoretical facts, infinite in number" (ibid., 135). A thermometer, for instance, can only tell us the temperature to a finite degree of accuracy. Thus, our approximate temperature measurement would translate mathematically into a interval of $\mathbb{R}$ with non-zero length. In phase space, our observations of some initial state would correspond to a smearing out of the initial conditions across some region with positive volume. We can reduce the size of this region by obtaining more precise measurements, but we can never reduce it to a point.

Duhem defines a criteria for when a mathematical deduction is useful for physicists. He writes that "A mathematical deduction... may therefore be useful or otiose, according to whether or not it permits us to derive a *practically definite* prediction of the result of an experiment whose conditions are *practically given*" (ibid., 137). In other words, the bundle of initial conditions corresponding to our initial measurement should not spread out too much over the system's phase space under time evolution. If we calculate each trajectory in the initial bundle to some future time, they should all (or at least mostly) be within some region that would correspond to another measurement of the system. If the calculated trajectories end up in many different regions of phase space, then this deduction is useless to the physicist.

---

[15] See §15.1 in Maz'ya and Shaposhnikova (1999) for an account of this.

[16] See Mawhin (1993) & Schmidt (2017) for notable exceptions.

Duhem recognizes that the uselessness of deductions is relative to our measuring capacities (ibid., 138). More accurate measurements would tighten the bundle of theoretical facts under examination, and this might lead to a formerly useless deduction becoming useful. However, he explicitly defines systems with sensitive dependence as being outside the domain of physics in a section titled "An Example of Mathematical Deduction That Can Never Be Utilized". He writes:

> But in vain do we tighten indefinitely the first bundle and make it as thin as possible; we are not authorized to diminish as much as we please the deviation of the second bundle; although the first bundle is infinitely narrow, the blades forming the second bundle diverge and separate out without our being able to reduce their mutual deviations below a certain limit. *Such a mathematical deduction is and always will remain useless to the physicist* (ibid, 139, emphasis added).

This is an apt description of what happens in chaotic systems. No matter how tightly packed an initial bundle of solutions is, it will eventually fan out over the chaotic region of phase space. Duhem believes that any deduction exhibiting this behavior will be forever useless to the physicist.

To exemplify the uselessness of such a mathematical deduction, Duhem references two systems. He first references Hadamard's system of a particle on a surface of negative curvature, found in Fig. 3 (ibid., 139–141). He then goes on to discuss Poincaré's work on the stability of the solar system (ibid, 142). Referencing Poincaré's demonstration of sensitive dependence in three-body systems, Duhem argues that although the question of stability may have meaning for the mathematician, it becomes "for the astronomer a question devoid of all meaning" (ibid, 142).[17] Same as Hadamard, Duhem rejects the idea that something observable, such as the solar system's stability, depends on unobservably precise initial conditions. Rather, questions about stability are meaningless for physicists.

Remarkably, Duhem anticipates that these systems, which we now call "chaotic", will prove to be widespread in the math of classical physics. He predicts that "a great many problems well defined for the mathematician [will] lose all their meaning for the physicist" (ibid., 141). Later on, he observes that "One cannot go through the numerous and difficult deductions of celestial mechanics and mathematical physics without suspecting that many of these deductions are condemned to eternal sterility" (ibid., 143). This last quote is particularly interesting because Duhem recognizes the relationship between the difficulty of calculating trajectories and the chaotic behavior he describes. This shows that Duhem is aware of the nature of chaos, and that it will occur frequently in the mathematics of classical physics, long before this knowledge reached the larger physics community. Here, it is Duhem's philosophical convictions that causes him to define physics in a way that excludes chaos.

In Duhem, we see the tension between chaos theory and positivism given vivid expression. First, he adeptly describes the mathematical behavior. Then, he bluntly rejects it as useless and meaningless on empirical grounds. Amongst the scrap heap of "useless deductions", we would no doubt find many of the interesting results described by modern chaos theory. Strange attractors and their fractal geometries, intersections of stable and unstable manifolds which grow infinitely tangled, global Lyapunov exponents, etc. — these are the invariants of motion for chaotic systems; some of which are already coming into view thanks to Poincaré and Hadamard. However, because they do not describe a correlation between observed initial and final states, they are to be thrown out as meaningless mathematical demonstrations of "false rigor" and safely ignored by the physicist.

---

[17] In the original French, "pour l'astronome, une question dénuée de tout sens" (Duhem, 1906, 230).

### 3.4. A standard view

We have just shown two prominent theorists of Poincaré's time — Jacques Hadamard and Pierre Duhem — displaying both a profound understanding of the mathematics of chaos while dismissing its physical significance on philosophical grounds. But how widespread were Hadamard and Duhem's views in the physics community? When we look back at Poincaré's original work on the three-body problem, we can see similar views reflected quite broadly.

While Poincaré was working on the three-body problem, astronomer Octave Callandreau writes him a letter of warning:

> First, in general, I can tell you that astronomers have not assigned the questions of stability the scope that geometers sometimes give it: They have desired to represent the motion of the elements of the solar system by taking into account the uncertainty of data gathered from observation; because a rigorous integration of the differential equations would, from an astronomical and practical point of view, be accompanied by the knowledge of fully precise data of the arbitrary mass elements (Callandreau, 1882).

Callandreau's reasoning, which he claims to be standard amongst astronomers, closely anticipates Hadamard and Duhem's. By taking into account the finite precision of observation, astronomers naturally confine their attention to situations where bundles of initial conditions will agree on future predictions. From this statement, we can infer that Hadamard and Duhem's critiques of chaos in dynamical systems were not contrarian at all. Rather, Hadamard and Duhem's arguments were explicit expressions of an existing empiricist blindspot to chaos.

In Poincaré's 1891 review article of his work on the three body problem, he plays defense to this empiricist critique. He tries to pre-empt the anticipated criticism that his findings will be "interesting only for the geometer and useless to the astronomer" (Poincaré, 2002, 88). He concedes that it is "useless to ask of more precision from the calculations than from the observations" (ibid., 86). However, Poincaré pleads with his audience to look past the sensitive dependence in his results. He directs their attention to his work on periodic trajectories, which he claims could be used as good first-order approximations for actual trajectories. Thus, even Poincaré knew that his work would be running counter to the empiricist currents of his day and struggled to articulate its physical significance.

### 3.5. Philosophy of science in the history of science

Through the influence of positivism in physics, we see philosophy of science guiding the history of science in many directions. The goal of positivists in physics was to discard the meaningless metaphysical baggage of physical theories. This could be done by examining how theory is grounded in experience. For the positivist, the laws of physics are ultimately just summaries of empirical regularities. Mathematical structure that strays too far beyond the regularities of experience is to be condemned as useless or meaningless. This conception of physics would simultaneously bring certain new insights into focus, while obscuring others.

Einstein would credit positivism as playing a key role in his discovery of relativity.[18] In a letter to Moritz Schlick, Einstein describes how he had studied David Hume and Ernst Mach before discovering relativity, writing that "It is very possible that without these philosophical studies I would not have arrived at the solution" (Einstein, 1998, 161). For this strange new theory, the strict empiricism of positivism proved to be a benefit for letting go of certain classical assumptions.

However, we have just seen that forays into what we now call chaos theory would be stifled by the quest for a metaphysics-free physics. By understanding physical theories to be nothing more than descriptions of observed regularities, positivism would smuggle in the assumption that physical systems are generically regular. Implicit in the writings of Hadamard and Duhem is the idea that the physical is delimited by the observable. The laws of physics were, by definition, observed regularities. It would simply make no sense to say that the laws themselves can give rise to a lack of observable correlations, or that unobservable differences create large observable ones.

---

[18] For an account of this, see Norton (2010).

Chaos did not fit into this Procrustean bed. Consider Smale's description about the development of chaos theory:

> Chaos developed not with the discovery of new physical laws, but by a deeper analysis of the equations underlying Newtonian physics. Chaos is a scientific revolution based on mathematics. Deduction rather than induction is the methodology. Chaos takes the equations of Newton, analyzes them with mathematics, and uses that analysis to establish the widespread unpredictability in the phenomena described by those equations (Smale, 2000, 8).

This reverses the positivist order of explanation. Newton's laws are not merely descriptions of observed regularities. They are the dynamics of real physical systems. Their application may or may not yield regular behavior. Thus, taking these laws seriously in their own right led to the novel insight that law-governed, deterministic systems can display highly uncorrelated behavior, or — in a word — chaos. It also allowed sensitive dependence to be viewed as a real physical phenomena, not a mathematical pathology. Thus, it is only after physicists and mathematicians started exploring the "useless" and "meaningless" questions of mathematical physics that the assumption of generic regular behavior could be rooted out.

As a caveat, nowhere am I claiming that chaos theory poses any problem to empiricism broadly construed. By now the existence and character of chaos has been observationally confirmed for many classical systems. Even some of its most surprising mathematical features, such as period doubling, have been observed in experiments (Libchaber et al., 1982). Nor do I think that sensitive dependence would be fully unintelligible to the positivist. After all, we can witness chaotic systems with *observably* small differences in initial states diverge exponentially. What I have argued is that, as a matter of history, positivist ideas about meaning, observation, and laws did obscure the significance of early results on chaos following Poincaré.[19]

## 4. Conclusion: Integrating the history

Given a positivist understanding of physical laws as well as a rejection of unobservable differences, Poincaré's contemporaries saw early results on chaos as mathematical curiosities of certain abstract classical systems, devoid of any profound implications for physics. To establish this claim, I have pointed to the writings of prominent mathematicians and physicists who worked near to Poincaré and had a clear understanding of the mathematics. How does this explanation fit into the larger history of chaos?

If positivist philosophy helped suppress chaos research, does the absence of positivist philosophy in the 1960s explain its resurgence? I would be cautious to make such a sweeping claim. It is true that after the quantum and relativity revolutions, particularly following the mathematical triumphs of Einstein in general relativity, physics had grown less leery of highly abstract mathematics.[20] But this period of chaos research was primarily driven by technological change, with philosophy of science bringing up the rear.

The resurgence of interest into chaotic systems was a convergence of two, initially independent threads. The first was initiated by Lorenz (1962, 1963), who was driven by numerical results. In this work, the many numerically generated graphs, phase portraits, and Poincaré sections did add an "empirical" aspect back into chaos research. They allowed the ideas of chaos theory to be more easily visualized and interpreted and less easy to dismiss. Still, many of these results were initially distrusted due to their numerical origins.[21]

The second thread was work in pure math done by figures such as Stephen Smale, Anosov (1967), and Sinai (1970). These results were obtained in the qualitative and topological fashion of Poincaré. When these two threads met, they provided mutual support. Rigorous results in mathematics gave researchers confidence in the numerical results.[22] Likewise, numerical techniques became essential to probing and verifying many of the mathematical conjectures of the field. This led to a shift in the philosophy of mathematics regarding the meaning of proof and evidence (Horgan, 1993). During this period, technological progress was the primary driver of change, with philosophy of science largely playing catch up.

A final point. I have argued that the positivist inclinations of Poincaré's *immediate* circle of contemporaries led to them dismiss chaos. But what about contemporaries outside of this immediate circle? Or researchers who followed shortly after? Positivist thinking was widespread, but it was by no means constant across space or time. How much blame can we place on positivism in suppressing these results across the entire intervening period?

Here is a different way to think about it. The time-frame for scientific ideas to have an impact is concentrated around when they are introduced. The specialists they are most likely to impact are concentrated around where they are introduced (especially in a period before rapid communication). Researchers with intimate familiarity with these results as they appeared — Hadamard, Duhem, and even Poincaré — were the ones who bore the responsibility of communicating their physical significance to the wider physics community. Given their philosophical commitments, they were unable to do so. Thus, the prevailing philosophy of science at the time led to these ideas being discarded as physically uninteresting, until eventually they were rehabilitated thanks to a chance encounter with Smale. And once something has been discarded, it is more remarkable when it is recovered than when it remains forgotten.

**Declaration of competing interest**

The authors declare that they have no known competing financial interests or personal relationships that could have appeared to influence the work reported in this paper.

**Acknowledgments**

I would like to thank John D. Norton, David Wallace, Michael Dietrich, Robert Batterman, Stephen Perry, Andrew Mugler, Tanner

---

[19] Interestingly, some of the most robust empirical predictions of chaos theory are, in a sense, meta-predictions. For example, a chaotic system's maximum Lyapunov exponent describes how fast nearby initial states diverge. In practice, this provides a second-order prediction for how quickly we can expect first-order predictions to break down.

[20] Following his work on general relativity, Einstein began to distance himself from the positivist ideas that led him to special relativity. He was particularly unsettled by their application in quantum theory by the likes of Heisenberg and Bohr (Frank, 1947, 216).

[21] Yoshisuke Ueda proceeded Lorenz's discovery of chaos using an analog computer in 1961. However, his advisor was unsure what to make of the results, so chose not to publish them (Ueda, 2000). Likewise, Mitchell Feigenbaum's numerical results on universality were initially dismissed as not meeting the standards of mathematical evidence (Gleick, 1987, ch. 6).

[22] Feigenbaum's ideas about universality were mathematically corroborated in Lanford III (1982). Mathematicians would also develop "shadowing lemmas", which showed that even though numerically generated trajectories are expected to diverge exponentially from their true trajectory, they nonetheless stay nearby, or "shadow", some true trajectory (Hammel et al., 1988). This allows us to infer that the detailed phase space structures exhibited by numerical solutions, such as strange attractors, are aspects of the real solutions and not just numerical artifacts.

*B. Park*     *Studies in History and Philosophy of Science 116 (2026) 102133*

## References

Anosov, D. V. (1967). Geodesic flows on closed Riemannian manifolds of negative curvature. *Trudy Matematicheskogo Instituta Steklov*, *90*, 3–210, URL http://mi.mathnet.ru/tm2795.

Anosov, D. V. (2006). Dynamical systems in the 1960s: The hyperbolic revolution. In †., A. A. Bolibruch, Y. S. Osipov, Y. G. Sinai, V. I. Arnold, A. A. Bolibruch, A. M. Vershik, Y. I. Manin, Y. S. Osipov, Y. Sinai, V. Tikhomirov, L. Faddeev, & V. B. Philippov (Eds.), *Mathematical events of the twentieth century* (pp. 1–17). Berlin, Heidelberg: Springer Berlin Heidelberg.

Ariew, R. (2022). Pierre duhem. In E. N. Zalta (Ed.), *The stanford encyclopedia of philosophy* (Spring 2022 ed.). Metaphysics Research Lab, Stanford University.

Arnold, V. I. (1963). Proof of a theorem of a.n. Kolmogorov on the preservation of conditionally periodic motions under a small perturbation of the Hamiltonian. *Uspekhi Matematicheskikh Nauk*, *18*(5).

Aubin, D., & Dalmedico, A. D. (2002). Writing the history of dynamical systems and chaos: Longue Durée and revolution, disciplines and cultures. *Historia Mathematica*, *29*(3), 273–339, URL https://www.sciencedirect.com/science/article/pii/S0315086002923517.

Barrow-Green, J. (1997). *History of mathematics*: vol. 11, *Poincaré and the three body problem*. Providence, RI: American Mathematical Society.

Birkhoff, G. D. (1913). Proof of Poincaré's geometric theorem. *Transactions of the American Mathematical Society*, *14*(1), 14–22.

Birkhoff, G. D. (1931). Proof of the ergodic theorem. *Proceedings of the National Academy of Sciences*, *17*(12), 656–660.

Blackmore, J. T. (1972). *Ernst Mach; his work, life, and influence*. Berkeley: University of California Press.

Callandreau, O. (1882). *3-9-2. Octave Callandreau to H. Poincaré*: *Henri poincaré papers*, Archives Henri Poincaré.

Comte, A. (1852/1909). *Catéchisme positiviste: ou, Sommaire exposition de la religion universelle, avec une introduction et des notes explicatives* (p. xxxvi+382). Paris: Garnier.

Comte, A. (1853). vol. 1, *The positive philosophy of auguste comte*. London: George Bell & Sons,

Comte, A. (1865). *A general view of positivism*. London: Trübner and Co..

Duhem, P. (1906). *La théorie physique: Son objet et sa structure*. Paris: Chevalier & Rivière.

Duhem, P. (1954). *The aim and structure of physical theory*. Princeton University Press.

Durkheim, É. (1895). *Les Règles de la méthode sociologique* (1st ed.). Paris: Félix Alcan.

Einstein, A. (1998). In R. Schulmann, A. J. Kox, M. Janssen, & J. Illy (Eds.), *The collected papers of Albert Einstein, Volume 8: The Berlin Years: Correspondence, 1914–1918*. Princeton, NJ: Princeton University Press.

Frank, P. (1947). In S. Kusaka (Ed.), *A borzoi book, Einstein: His life and times* (p. 298). New York: A. A. Knopf, URL https://books.google.com/books?id=Y9DaAAAAMAAJ.

Franks, J. (1989). Review of James Gleick, 1987. Chaos. Making a new science, viking penguin, new york 1987. *The Mathematical Intelligencer*, *11*(1), 65–69.

Gleick, J. (1987). *Chaos: Making a new science*. Viking Books.

Hadamard, J. (1898). Les surfaces à courbures opposées et leurs lignes géodésique. *Journal de Mathématiques Pures et Appliquées*, *4*, 27–73.

Hadamard, J. (1901). *Notice sur les travaux scientifiques*. Gauthier-Villars.

Hadamard, J. (1923). *Lectures on Cauchy's problem in linear partial differential equations*. New Haven: Yale University Press.

Hadamard, J. (1927). L'oeuvre de Duhem dans son aspect mathématique. *Mémoires de la Société des Sciences Physiques et Naturelles de Bordeaux*, *1*, 635–665.

Hammel, S. M., Yorke, J. A., & Grebogi, C. (1988). Numerical orbits of chaotic processes represent true orbits. *Bulletin (New Series) of the American Mathematical Society*, *19*(2), 465–469.

Hilborn, R. C. (2000). *Chaos and nonlinear dynamics: an introduction for scientists and engineers* (2nd ed.). Oxford: Oxford University Press.

Hirsch, M. W. (1989). Chaos, rigor, and hype. *The Mathematical Intelligencer*, *11*(3), 6–13.

Hofstadter, D. (1985). *Metamagical themas: questing for the essence of mind and pattern*. Basic Books.

Horgan, J. (1993). The death of proof. *Scientific American*, *269*(4), 92.

Hume, D. (1748). *Oxford world's classics, An enquiry concerning human understanding*. Oxford University Press, 2008.

(1900). JEUDI, 2 aout: Séances de section. *Revue de Métaphysique et de Morale*, *8*(5), 525–565.

Kellert, S. H. (1993). *In the wake of chaos: Unpredictable order in dynamical systems*. Chicago: University of Chicago Press.

Kolmogorov, A. N. (1954). On the conservation of conditionally periodic motions under small perturbation of the Hamiltonian. *Doklady Akademii Nauk SSSR*, *98*(4), 527–530.

Lanford III, O. E. (1982). A Computer-Assisted proof of the Feigenbaum conjectures. *Bulletin (New Series) of the American Mathematical Society*, *6*(3), 427–434.

Libchaber, A., Laroche, C., & Fauve, S. (1982). Period doubling cascade in mercury, a quantitative measurement. *Journal de Physique Lettres*, *43*(7), 211–216.

Lorenz, E. N. (1962). The statistical prediction of solutions of dynamical equations. In *Proceedings of the international symposium on numerical weather prediction, 1962* (pp. 629–635). Tokyo, Japan: Meteorological Society of Japan.

Lorenz, E. N. (1963). Deterministic nonperiodic flow. *Journal of the Atmospheric Sciences*, *20*(2), 130–141.

Lorenz, E. N. (1969). The predictability of a flow which possesses many scales of motion. *Tellus*, *21*(3), 289–307.

Mach, E. (1893). *The science of mechanics*. Chicago: Open Court Publishing.

Mawhin, J. (1993). *The Centennial Legacy of Poincaré and Lyapunov in Ordinary Differential Equaitons*. Institut de mathématique pure et appliquée, Université catholique de Louvain.

Maxwell, J. C. (1877/1920). *Matter and motion*. New York: The MacMillan Co..

Maz'ya, V. G., & Shaposhnikova, T. O. (1999). *History of mathematics*: vol. 14, *Jacques Hadamard: a universal mathematician*. Providence, RI: American Mathematical Society.

Mill, J. S. (1866). *Auguste comte and positivism* (p. 210). Paternoster Row, London: N. Trübner & Co..

Moser, J. (1962). On invariant curves of area-preserving mappings of an annulus. *Nachrichten der Akademie der Wissenschaften in Göttingen Mathematisch-Physikalische Klasse II*, 1–20.

Newton, I. (1704). *Opticks: or, A treatise of the reflexions, refractions, inflexions and colours of light*. London: Sam Smith and Benj. Walford.

Norton, J. D. (2010). How Hume and Mach helped Einstein find special relativity. In M. Dickson, M. Domski, & M. Friedman (Eds.), *Discourse on a new method: reinvigorating the marriage of history and philosophy of science* (pp. 359–386). La Salle, IL: Open Court.

Palmer, T. N., Döring, A., & Seregin, G. (2014). The real butterfly effect. *Nonlinearity*, *27*(9), R123–R141.

Peterson, I. (1993). *Newton's clock: Chaos in the solar system*. New York: W. H. Freeman and Company.

Poincaré, H. (1890). Poincaré à Hertz, 1890-11-29. *Archives Henri Poincaré*, URL http://henripoincare.fr/s/correspondance/item/11871.

Poincaré, H. (1905). *Science and hypothesis*. New York: The Walter Scott Publishing Co..

Poincaré, H. (1914). *Science and method*. London: Thomas Nelson and Sons.

Poincaré, H. (1967a). *New methods of celestial mechanics*: vol. 1, National Aeronautics and Space Administration.

Poincaré, H. (1967b). *New methods of celestial mechanics*: vol. 2, National Aeronautics and Space Administration.

Poincaré, H. (1967c). *New methods of celestial mechanics*: vol. 3, National Aeronautics and Space Administration.

Poincaré, H. (2002). *Scientific opportunism l'opportunisme scientifique: An anthology*. Birkhäuser.

Poincaré, H. (2017). *The three-body problem and the equations of dynamics: Poincaré's foundational work on dynamical systems theory*: vol. 443, Springer.

Ruelle, D. (1993). *Chance and chaos*: vol. 110, Princeton University Press.

Schmidt, J. C. (2017). Science in an unstable world. On Pierre Duhem's challenge to the methodology of exact sciences. In *Berechenbarkeit der Welt?* (pp. 403–434). Springer.

Sinai, Y. G. (1970). Dynamical systems with elastic reflections. *Russian Mathematical Surveys*, *25*(2), 137.

Smale, S. (1967). Differentiable dynamical systems. *Bulletin of the American Mathematical Society*, *73*(6), 747–817.

Smale, S. (1998). Finding a horseshoe on the beaches of Rio. *The Mathematical Intelligencer*, *20*(1), 39–44.

Smale, S. (2000). On how I got started in dynamical systems 1959–1962. In *The chaos avant-garde: memories of the early days of chaos theory* (pp. 1–6). World Scientific.

Steiner, F. (1994). Quantum chaos. *Path Integrals from PeV To TeV*, 12.

Ueda, Y. (2000). Strange attractors and the origin of chaos. In R. Abraham, & Y. Ueda (Eds.), *The chaos avant-garde: memories of the early days of chaos theory* (pp. 23–55). World Scientific.

Werndl, C. (2009). What are the new implications of chaos for unpredictability? *The British Journal for the Philosophy of Science*, *60*(1), 195–220.

Zuchowski, L. C. (2014). Gestalt switches in poincaré's prize paper: An inspiration for, but not an instance of, chaos. *Studies in History and Philosophy of Science. Part B. Studies in History and Philosophy of Modern Physics*, *47*(1), 1–14.